
%
%
%
%
%

\input phyzzx

\sequentialequations

\overfullrule=0pt
\catcode`\@=11
\def\mm{matrix model}
\def\ns{non-singlet}

\def \DM{ {\partial \over {\partial \mu}}}

\def\NP{{\it Nucl. Phys.\ }}

\def\PL{{\it Phys. Lett.\ }}
\def\PR{{\it Phys. Rev.\ }}
\def\PRL{{\it Phys. Rev. Lett.\ }}

\def\JTP{{\it JETP \ }}
\def\JP{{\it J. Phys.\ }}
\def\IJMP{{\it Int. Jour. Mod. Phys.\ }}
\def\Mod{{\it Mod. Phys. Lett.\ }}

\def\td{two-dimensional}
\def\lc{light cone}
\def\KT{Kosterlitz-Thouless}
\def\eqaligntwo#1{\null\,\vcenter{\openup\jot\m@th
\ialign{\strut\hfil
$\displaystyle{##}$&$\displaystyle{{}##}$&$\displaystyle{{}##}$\hfil
\crcr#1\crcr}}\,}
\catcode`\@=12

\REF\GM{D.~J.~Gross and A.~A.~Migdal, \PRL {\bf 64} (1990) 717;
M. Douglas and S.~Shenker, \NP {\bf B335} (1990) 635;
E.~Brezin and V.~Kazakov, \PL {\bf 236B} (1990) 144.}
\REF\Doug{ M. Douglas, \PL {\bf 238B} (1990) 176.}
\REF\GMil{D. J. Gross and N. Miljkovic \journal Phys. Lett.
& 238B (1990) 217; E. Brezin, V. Kazakov, and Al. B. Zamolodchikov
\journal Nucl. Phys. &B338 (1990) 673; P. Ginsparg and J. Zinn-Justin
\journal Phys. Lett. &240B (1990) 333; G. Parisi \journal
Phys. Lett. &238B (1990) 209.}
\REF\GKl{D. J. Gross and I. R. Klebanov, \NP {\bf B344} (1990) 475.}
\REF\Polch{J. Polchinski \journal Nucl. Phys. &B346 (1990) 253. }
\REF\CThorn{C. B. Thorn, \PL {\bf 70B} (1977) 85; \PR {\bf D32} (1978) 1073.}
\REF\Brod{H.-C. Pauli and S. Brodsky, \PR {\bf D32} (1985) 1993, 2001;
K. Hornbostel, S. Brodsky and H.-C. Pauli, \PR {\bf D41} (1990) 3814;
for a good review, see K. Hornbostel, Ph. D. thesis, SLAC report
No. 333 (1988).}
\REF\KS{I. R. Klebanov and L. Susskind, \NP {\bf B309}, 175 (1988).}
\REF\DM{S. Dalley and T.R. Morris, \IJMP {\bf A5}, 3929 (1990).}
\REF\ind{S. Das, S. Naik and S. Wadia, \Mod {\bf A4} (1989) 1033;
J. Polchinski, \NP {\bf B324 } (1989) 123;
S.~Das, A.~Dhar and S. Wadia, \Mod {\bf A5} (1990) 799;
T. Banks and J. Lykken, \NP {\bf B331} (1990) 173;
A. Tseytlin, \IJMP {\bf A5} (1990) 1833.}
\REF\CT{T. L. Curtright and C. B. Thorn, \PRL {\bf 48}, 1309 (1982).}
\REF\Thorn{C. B. Thorn, \PL {\bf B242}, 364 (1990).}
\REF\Vort{D. J. Gross and I. R. Klebanov
 \journal Nucl. Phys. &B354 (1991) 459.}
\REF\bk{D. Boulatov and V. Kazakov, LPTENS preprint 91/24.}
\REF\bkt{V. L. Berezinskii, \JTP {\bf 34} (1972) 610;
M. Kosterlitz and D. Thouless, \JP {\bf C6} (1973) 1181.}
\REF\Wein{D. Weingarten, \PL {\bf 90} (1980) 280.}
\REF\Dall{S. Dalley, \Mod {\bf A7}, 1651 (1992).}

\def\eqaligntwo#1{\null\,\vcenter{\openup\jot\m@th
\ialign{\strut\hfil
$\displaystyle{##}$&$\displaystyle{{}##}$&$\displaystyle{{}##}$\hfil
\crcr#1\crcr}}\,}
\catcode`\@=12

\def\qg{quantum gravity}

\def\half{{1\over 2}}
\def\d{\dagger}

\nopagenumbers

{\baselineskip=12pt
\line{\hfil PUPT-1333}
\line{\hfil hepth@xxx/9207065}
}
\title{Light Cone Quantization of the $c=2$ Matrix Model}
\author{Simon Dalley and Igor R. Klebanov }
\JHL
\abstract
We study the large $N$ limit of an
interacting \td\ matrix field theory, whose perturbative expansion
generates the sum over planar random graphs embedded in two dimensions.
In the \lc\ quantization the theory possesses closed string excitations
which become free as $N\to\infty$. If the longitudinal momenta
are discretized, then the calculation of the free string spectrum
reduces to finite matrix diagonalization, the size of the matrix
growing as the cut-off is removed. Our numerical results suggest
that, for a critical coupling, the \lc\ string spectrum
becomes continuous. This would indicate the massless dynamics
of the Liouville mode of \td\ gravity, which would constitute
a {\it third} dimension of the string theory.
\endpage

\pagenumbers
\vsize=8.9in
\hsize=6.5in
\centerline{\caps 1. Introduction}
\bigskip

Recently there has been  considerable renewed interest in
large-$N$ \mm s. The one-dimensional hermitian matrix chain models
have been solved in the double-scaling limit and identified with
the $c<1$ minimal models coupled to \td\ \qg [\GM, \Doug].
In the same fashion
the hermitian matrix quantum mechanics, which describes the $c=1$ theory,
has been solved both for non-compact [\GMil] and for circular
target space [\GKl]. The $c=1$
model can be interpreted in terms of  $D=2$ string theory [\Polch], where
the role of the extra dimension is played by the conformal factor
of the world sheet \qg. One should keep in mind that the models
with $c>1$ are of   much greater interest because they are expected
to correspond to string theories in $D>2$. These theories should
have a much richer structure than in $D=2$ because strings
can exhibit transverse oscillations. Unfortunately, although the
$c\leq 1$ models have been solved exactly to all orders in the genus
expansion, little is known about the $c>1$ theories.

In this paper we begin to fill this gap by studying a $c=2$ \mm,
which is defined in the euclidean space as a \td\ field theory
with the action
$$ S=\int d^2 x \Tr \left (\half (\partial_\alpha M)^2+\half \mu M^2-
{1\over 3\sqrt{N}}\lambda M^3 \right )\ ,
\eqn\action$$
where $M(x^0, x^1)$ is an $N\times N$ hermitian matrix field.
The connection of this \mm\ with triangulated \td\ \qg\ follows,
as usual, after identifying the Feynman graphs with the graphs
dual to triangulations. The lattice link factor is the two-dimensional
scalar propagator,
$$ G(\vec x_i, \vec x_j)= \int {d^2 p \over (2\pi)^2 }{{\rm e}^{i
\vec p \cdot (\vec x_i -\vec x_j) }
\over p^2+\mu} = {1\over 2\pi} K_0 \bigl(\sqrt\mu~ |\vec x_i-\vec x_j|
\bigr)
\ ,\eqn\eq$$
where $K_0$ is a modified Bessel function.
Thus, at the leading
order in $N$, we obtain
a sum over the planar triangulated random surfaces
embedded in two dimensions.\foot{Note that, as for $c=1$, the
coordinates are specified at the centers of the triangles.}
The logarithmic divergence of $G$ at small separations is very mild.
If the tadpole graphs are discarded, as they should be because they
do not correspond to good triangulations, then the entire perturbative
expansion of the theory \action\ is finite.
This is similar to what we find in the matrix models for $c\leq 1$.
Therefore, it is sensible to look for a singularity in the sum
over the planar graphs as a function of the dimensionless parameter
$\lambda/\mu$. The primary purpose of this paper is to look for
such critical behaviour, and to discuss the non-critical string theory
that results.

The arguments above suggest that, although the $c=2$ \mm\ of eq.
\action\ is certainly more complex than the matrix quantum mechanics,
we should still be able to take advantage of the simplifications
that distinguish the \td\ field theories. The method to pursue this
that we believe is particularly convenient, is to continue $x^0\to ix^0$
and to carry out \lc\ quantization of the resulting
$(1+1)$-dimensional field theory. This is often the most efficient way
to study a theory. Its additional advantage here is that, in the
large-$N$ limit, one can formulate a {\it linear} \lc\ Schroedinger
equation satisfied by free string states. After a cut-off is introduced
in the form of discretized longitudinal momentum [\CThorn, \Brod],
this equation
can be solved for the free string spectrum. Subsequently, this spectrum
can be studied as the cut-off is being removed.

The procedure outlined above was used in a paper by Susskind and one of
the authors [\KS] to study a $c>2$ \lc\ lattice gauge theory of a
large-$N$ complex matrix. It was shown that this theory can be driven
into a phase where the longitudinal momentum of each string bit
assumes the smallest possible value. In this phase, each of the
$c-2$ transverse dimensions becomes a free massless scalar field
in the parameter space of the string. Then one recovers
the free critical string spectrum [\KS, \DM] without any necessity of
taking the transverse lattice spacing to zero.

We could look
for this ``critical string phase'' in the model of eq. \action,
but the result would turn out uninteresting, as the theory has $c=2$
and would end up with no degrees of freedom.
In this paper we instead look for a totally different phase of \lc\
large-$N$ matrix models. In this phase, the longitudinal momentum
of each string bit is not frozen at its minimum value, but is
instead allowed to fluctuate. Therefore, the \lc\ dynamics
of the theory \action\ is quite non-trivial, being described by fluctuations
in the number of string bits  and their longitudinal momentum.
We will formulate the details of these dynamics in section 2.
Our intuitive expectation is that, as $\lambda/\mu$ is driven
to its critical value, these fluctuations implement the effects
of the conformal factor of the continuum
\td\ \qg, producing a continuous spectrum
of the \lc\ hamiltonian. Then the model represents a $D=3$ string
theory. This would be an explicit demonstration that the rule $D=c+1$
[\ind] applies to non-critical string theory with $c>1$.

It is appropriate to call the phase we are after the ``non-critical
string phase'' of the theory \action. Our study of this phase
will be based on exact numerical diagonalizations of the
\lc\ hamiltonian with discretized longitudinal momenta. This is
a convenient cut-off because it renders the total number of
single-string states finite. However, the number of states
grows rapidly as the cut-off is being removed, and the
calculations we have
managed to carry out are rather far from the continuum limit.
We cannot yet draw any conclusions,
but the results that we present in section 3 do encourage us to
believe that the theory has a critical point where the spectrum is
continuous due to the Liouville mode.
\bigskip

\centerline{\caps 2. Light Cone Quantisation}
\bigskip

In \lc\ quantisation we treat $x^{+}=(x^0+x^1)/\sqrt 2$
as the time variable, and from the
Minkowskian version of the action $S$ in eq. \action\ derive;
$$ \eqalign{&P^{+}(x^{+}) = \int dx^{-} \Tr (\partial_{-}M)^{2}\ ,\cr
&P^{-}(x^{+}) = \int dx^{-} \Tr (\half \mu M^2 - {\lambda \over 3\sqrt{N}}
M^3 )\ .\cr }\eqn\pminus$$
The canonical commutation relations are imposed at equal $x^{+}$ lines;
$$ [M_{ij}(x^{-}),\partial_{-}M_{kl}(\tilde{x}^{-})] = {i\over 2}
\delta (x^{-} -
\tilde{x}^{-})\delta_{il} \delta_{jk}\ .\eqn\rel$$
We may expand in Fourier modes\foot{The symbol $\dagger$ is always understood
to have purely quantum meaning and never acts on indices.};
$$M_{ij}={1 \over \sqrt{2\pi}} \int_{0}^{\infty} {dk^{+} \over \sqrt{2k^{+}}}
(a_{ij}(k^{+}){\rm e}^{-ik^{+}x^{-}} + a_{ji}^{\d}(k^{+})
{\rm e}^{ik^{+}x^{-}})\eqn\Four$$
to obtain
$$[a_{ij}(k^{+}),a_{lk}^{\d}(\tilde{k}^{+})] = \delta(k^{+} - \tilde{k}^{+})
\delta_{il}\delta_{jk}\ .\eqn\modeccr$$
After substituting eq. \Four\ into eq. \pminus, we obtain
$$\eqalign{
&:P^+: = \int_{0}^{\infty} dk^+ k^+ a_{ij}^{\d}(k^+)a_{ij}(k^+)\ ,\cr
&:P^{-}: =
\half \mu \int_{0}^{\infty} {dk^{+}\over k^{+}} a_{ij}^{\d}(k^{+})
a_{ij}(k^{+}) -{\lambda \over 4\sqrt{N\pi}}\cr
&\times\int_{0}^{\infty}
{dk_{1}^{+}dk_{2}^{+}\over \sqrt{k_{1}^{+}k_{2}^{+}(k_{1}^{+} + k_{2}^{+})}}
\left\{a_{ij}^{\d}(k_1^+ +k_2^+)a_{ik}(k_2^+)a_{kj}(k_1^+) +
a_{ik}^{\d}(k_1^+)a_{kj}^{\d}(k_2^+)a_{ij}(k_1^+ +k_{2}^+)\right\}
\cr }\eqn\nominus$$
(repeated indices are summed over). The normal
ordering is equivalent to removing the tadpole graphs in the Lagrangian
approach.
The states of the Fock
space are of the form;
$$a_{i_{1}j_{1}}^{\d}(k_{1}^{+})\cdots a_{i_{B}j_{B}}^{\d}(k_{B}^{+})
|0>\eqn\Fock$$
and the vacuum $|0>$ is an eigenstate of the fully interacting \lc\ Hamiltonian
$:P^{-}:$.

The key simplifying feature of \lc\ quantisation is that positive
energy quanta require $k^{+}>0$. In this way if the
allowed $k^{+}$ are discrete and non-zero, then one can enumerate
all the states of some given total longitudinal momentum $P^{+}$
[\CThorn, \Brod].
In order
to implement the discretisation we shall compactify the $x^{-}$ direction
to a circle of length $L$, so that the matrix field is periodic,
$M_{ij}(x^-)=M_{ij}(x^- +L)$. The allowed values of $k_b^+$ are
then $2\pi n_b/L$.
Note also from \nominus\ that quanta of $k^{+}=0$ are at infinite
energy if $\mu >0$ and so the integers $n_b$ are restricted to be
positive. For $P^+=2\pi K/L$, we find
$\sum_{b=1}^{B} n_{b}= K$.
The positive integer $K$ plays the role of a cut-off, and sending
it to infinity corresponds to removing the cut-off in the discretized
\lc\ quantization [\Brod].

Before proceeding we need to elaborate upon the \lc\ string picture of our
analysis. In \mm s the states in \Fock\ which have an
unambiguous description as single closed strings are of the form;
$$N^{-B/2} \Tr [a^{\d}(k_{1}^{+})\cdots a^{\d}(k_{B}^{+})] |0>
\ ,\eqn\string$$
in other words the singlet states under the $M\to \Omega^{\dagger} M\Omega$
global $SU(N)$ symmetry of  the action $S$.
This represents a boundary of length $B$ in the
dynamical triangulation. Each $a^{\d}(k^{+})$ creates a string bit carrying
longitudinal momentum $k^{+}$. It is important to realise that there is no
particular reason to believe that the non-singlet states, which have no simple
string identification, are unimportant as the cutoff is removed; they may well
have energies of the same order as those of the states \string\ . This may be
a crucial flaw in such a matrix model as we are considering here, to which
we will return in section 4.

In the limit $N \to \infty$ the \lc\ Hamiltonian
$P^{-}$ takes single closed string states to single closed string states.
One can easily check that the terms that convert one closed string into
two closed strings (two oscillator traces acting on the vacuum) are
of order $1/N$. Thus, as expected, the string coupling constant is
$\sim 1/N$, and sending it to zero
allows us to study the spectrum of free closed string states.
Because $P^{-}$ is
$SU(N)$ invariant it does not couple these closed strings to the non-singlet
states and we shall ignore the latter sector  of the theory. A more fundamental
reason for taking $1/N \to 0$ is that one expects simple bosonic string
 theories at $c>1$ to be tachyonic, so while there is no objection to studying
the free theory, the vacuum would collapse if we did not turn off the
interactions.

Now in the \lc\ formalism of critical string theory the longitudinal momentum
supported between two points on the string is proportional to the amount of
$\sigma$-space between these points. We can adopt a similar co-ordinate
system for the non-critical strings \string\ . Indeed, fixing a particular
bit as origin, we can define a positive scalar field on this $\sigma$-space
by $X=\Delta b/\Delta \sigma$, where $b$ is the  distance,
measured in number of bits, from the origin. As we remove
the cutoff on the longitudinal momentum allowed for bits, and hence on
discreteness of $\sigma$-space,
the scalar field will generically take constant values almost
everywhere in $\sigma$-space. We would like to be able to tune the theory
to a critical point where the scalar field is in a long wavelength regime.
Then it would be natural to identify it as an effective extra dimension in
which the string fluctuates, an interpretation commonly given to the Liouville
mode. In the next section we will attempt to identify where such a critical
point might exist. Of course it may not exist, but there is some hope that
an extra massless field arises since there is a known example of $c=2$ free
string theory where the spectrum is basically that of a single free
massless field. This
picture emerges both from the naive extension of Liouville theory to $c>1$
[\CT] and
a naive extension of dual string theory to $c<25$ [\Thorn].
The relationship between
such a free field and the scalar defined above would presumably be quite
complicated however.
\bigskip

\centerline{\caps 3. Exact Diagonalizations and Numerical Results}
\bigskip

Having discretised the longitudinal momenta $k^{+} = nP^{+}/K$, we
define new discrete mode expansions
$$M_{ij}={1 \over \sqrt{4\pi}} \sum_{n=1}^\infty {1\over \sqrt{n}}
\left (A_{ij}(n){\rm e}^{-iP^{+}nx^-/K} + A_{ji}^{\d}(n)
{\rm e}^{iP^{+}nx^-/K} \right )\ .\eqn\dm$$
The oscillator algebra is
$$[A_{ij}(n),A_{lk}^{\d}(n')] = \delta_{n n'}
\delta_{il}\delta_{jk}\ .\eqn\modecomm$$
A normalized closed string state of $B$ bits is of the form
$${1\over N^{B/2}\sqrt{s}}
\Tr [A^{\d}(n_1)\cdots A^{\d}(n_B)] |0>\ .\eqn\dstring$$
The states are defined by ordered partitions of
$K$ into $B$ positive integers, modulo cyclic permutations.
Therefore, the closed strings are oriented.
If $(n_1,~ n_2,~ \ldots,~ n_B)$ is taken into itself
by $s$ out of $B$ possible cyclic permutations,
then the corresponding state receives a
normalisation factor $1/\sqrt{s}$. In the absence of special
symmetries, $s=1$.
The \lc\ hamiltonian for strings of momentum $P^{+}$ becomes
$$:P^{-}: = {K\over 2P^{+}}\left(\mu V -{\lambda\over 2\sqrt\pi} T\right)
\ .\eqn\ham$$
The potential term;
$$V=\sum_{n=1}^{\infty} {1\over n}
A^{\d}_{ij}(n)A_{ij}(n) \ ,\eqn\pot$$
measures the tensional energy of a string without changing its state.
The kinetic term;
$$T={1\over\sqrt N}\sum_{n_1,n_2=1}^{\infty}
{A^{\d}_{ij}(n_1+n_2)A_{ik}(n_2)A_{kj}(n_1)
+A^{\d}_{ik}(n_1)A^{\d}_{kj}(n_2)A_{ij}(n_1+n_2)
\over \sqrt{n_1 n_2(n_1+n_2)}}
\ ,\eqn\kin$$
changes string states by
splitting bits into two and joining adjacent bits. Let us rewrite eq.
\ham\ as
$${2P^+P^-\over\mu} = K\left(V -x T\right) \,
\eqn\newform$$
where $x=\lambda/(2\mu\sqrt\pi)$ is the dimensionless parameter of the problem.
Compare eq. \newform\ with the usual form
for the  \lc\ string dispersion relation,
$$2P^+P^-\alpha' = P_\perp^2\alpha'+4 r-{D-2\over 6} \ ,
\eqn\eq$$
where $r$ is a non-negative integer. $\mu$ is the quantity of dimension
${\rm mass}^2$ which in our approach plays the role of string tension
$1/\alpha'$. The most basic evidence for the presence of transverse
dimensions is the continuous spectrum of the operator $2P^+P^-/\mu$.
For a finite cut-off $K$, it is a finite dimensional symmetric real
matrix whose spectrum is necessarily discrete. We can, however, look
for the critical value of $x$ where the spectrum becomes dense in the
$K\to\infty$ limit. In practice, we work with finite $K$ and look for
increasing density of eigenvalues.

To illustrate the calculation with a simple example, consider $K=3$
where there are only three closed string states.
Working to leading order in $N$, the normalized states are
$$ \eqalign{&|1>=
{1\over N^{3/2}\sqrt 3}\Tr [A^\dagger (1) A^\dagger(1)A^\dagger(1)]
|0>\ ,\qquad\qquad
|2>={1\over N}\Tr [A^\dagger (2) A^\dagger(1)] |0>\ ,\cr
&|3>={1\over N^{1/2}}\Tr [A^\dagger (3)] |0>\ .
\cr }\eqn\eq$$
A quick calculation gives
$$ K(V-xT)=\left (\matrix { 9 & -3\sqrt {3\over 2}~ x & 0\cr
-3\sqrt {3\over 2}~ x & {9\over 2} & -\sqrt 6~ x \cr
0 & -\sqrt 6~ x & 1\cr } \right )
\ . \eqn\eq$$

As $K$ increases, the number of states grows rapidly.
We used {\it Mathematica} to generate the
states, evaluate eq. \newform\
and diagonalise it over the range $0<x<1$. Figure~1
shows the eigenvalues at $K=7$ (19 states, the lowest 16 of which are shown)
as a function of $x$. This illustrates the general qualitative features.
The number of states is always odd.
$V$ is a diagonal matrix with positive entries;
$T$ is
a (relatively sparse) symmetric matrix with zeros on the
diagonal, having eigenvalues symmetrically distributed about zero. As $x$ is
increased from zero, many eigenvalues begin to decrease in such a way
that the density of low-lying eigenvalues increases.
\foot{The few eigenvalues that monotonically increase with
$x$ are presumably artifacts of the discretized problem that decouple
in the continuum limit.} On the other hand, for large $x$ the spectrum
is approximately that of the kinetic term multiplied by $x$, and the
gaps become
large. Therefore, there seems to be a possibility of fine tuning
$x$ so that the potential and the kinetic term are carefully balanced
to produce a denser and denser spectrum as $K$ increases.
To search for such behavior, we plotted
the mean square separation between adjacent
energy levels for the lower $50\%$ of the states.
This is shown in Figure~2 (solid curves).
The dashed curve shows the lowest eigenvalue in each case.
These plots clearly show a maximum density of eigenvalues around $x_{c} \sim
0.64 - 0.7$,
with a slight downward drift of $x_{c}$ with increasing $K$.
Since the solid curves of Figure~2 are quite bumpy due to many levels
crossing we also plotted the difference between the lowest two eigenvalues,
shown in Figure~3 for $K=8,10,12$. This demonstrates more dramatically the
decrease in level separation around $x_c$.
Although the approach to any sort of critical regime seems to be slow, we
 must keep in mind the low cut-off employed. Moreover one is reminded that at
$c=1$ the approach to criticality was logarithmic. If the minima
in Fig. 2 are
indicating an approach to a continuous density of eigenvalues, then it
is significant that the  ground state seems consistently tachyonic at that
point, as one expects for any $D>2$
string theory.

Clearly, at the moment the
data is at best suggestive and calculations with higher $K$ are needed.
A clear advantage of the discretized \lc\ quantization is that
it renders the problem perfectly regularized and numerically tractable.
A disadvantage is in the rapid growth of the size of the matrices
to be diagonalized. However, if we are only interested in
the low-lying spectrum, exact diagonalizations may be replaced
by approximate techniques which could allow significantly larger
$K$ to be probed.
\bigskip

\centerline{\caps 4. Discussion}
\bigskip

In the preceding sections we have presented some, albeit not
entirely conclusive, evidence that the \lc\ singlet spectrum
of the infinite $N$ \mm\ \action\ is continuous. Thus, the
model may give us useful information about a free $D=3$ string
theory. If this conjecture is confirmed by more detailed numerical
studies, is there any hope that, for a finite $N$ eq. \action\
defines a consistent interacting $D=3$ string theory?
Unfortunately, we believe that the answer is negative.

One problem with this theory is that, on general grounds, it is expected
to be tachyonic. Our numerical results appear to agree with that,
because at the coupling strength where the spectrum has the highest
density there is always a negative eigenvalue of $P^-$.
Another problem, which is perhaps even more severe, comes from the
states that transform non-trivially under $SU(N)$, such as the
adjoint representation states of the form
$$ N^{(1-B)/2} A^\dagger_{ij}(n_1)
A^\dagger_{jk}(n_2) \ldots
A^\dagger_{rs}(n_{B-1})
A^\dagger_{st}(n_{B}) |0>
\ . \eqn\eq$$
As discussed previously, the $SU(N)$ \ns\ states
cannot be identified with closed strings. One indication of that
is their diverging degeneracy factors in the $N\to\infty$ limit.
Perhaps the \ns s can be thought of as  closed strings that
have disintegrated into separate bits. Of course, one way of preventing
this is through a confinement phenomenon which would push their
energies to infinity in the continuum limit. In fact, confinement
is at work in the $c=1$ string theory, where the \ns\ energies diverge
logarithmically in the cut-off [\Vort, \bk]. The phase transition
at a critical value of the target space radius, which was
observed in the matrix model [\GKl], corresponds physically
to the \KT\ deconfinement of the world sheet vortices [\bkt].

Could the $c=2$ model of eq. \action\ also expel the \ns\ states to
infinite energy? To address this question, we have studied the
action of the \lc\ hamiltonian on the states in the adjoint
representation of $SU(N)$. Once again, we find a Schroedinger
equation in the $N\to \infty$ limit. We calculated its energy
levels exactly for low values of the cut-off $K$.
We find that the lowest eigenvalue of $2P^+ P^-/\mu$ starts
at 1 for $x=0$ and decreases monotonically with increasing $x$.
Although it is always higher than the lowest singlet eigenvalue,
it is impossible for this gap to diverge as $K\to\infty$.
In fact, a variational upper bound of 1 easily follows for the
lowest eigenvalue, if we
consider $A_{ij}^\dagger (K)|0>$ as the variational state.
Therefore, there is no confinement, and the \ns\ states
probably cause additional problems for the interacting string
theory. Some heuristic arguments with similar conclusions
were made in the context of $c>1$ models with discrete target
spaces [\Vort]. The proliferation of the \ns s was associated with
the \KT\ vortices which wind around the target space plaquettes
of lattice size. An advantage of the \lc\ approach is that
the \ns\ spectrum can be found with the same technique as
the singlet spectrum. It would be nice to carry out
a more detailed numerical study of the spectrum in the adjoint
representation.

The lack of confinement of the \ns\ states can be traced
to the fact that the action \action\ has only the global
$SU(N)$ symmetry, which is not gauged.
Gauging the $SU(N)$ will certainly lead to the confinement,
and the non-singlets will be pushed to infinite energy.
The effect of gauging the $SU(N)$ symmetry on the \td\ theory of
eq. \action\ is  to give the action
$$ S_{gauged}=\int d^2 x \Tr \left ({1\over 4g^2}F_{\alpha\beta}^2+
\half (\partial_\alpha M+i[A_\alpha, M])^2+\half \mu M^2-
{1\over 3}{\lambda\over\sqrt N} M^3 \right )
\eqn\nac$$
This theory may have a better chance than \action\ of describing
a sensible $c=2$ non-critical string. Actually, eq. \nac\ is somewhat related
to the well-known Weingarten model [\Wein], formulated for non-critical strings
[\Dall]. This issue, as
well as the possibility of a \lc\ solution of \nac\ will be
left for the future work.

In conclusion, we believe that the \lc\ quantization has
provided us with some new insights into the structure of
large-$N$ matrix models and of non-critical string theories with
$c>1$. Further progress may be achieved through numerical
solutions of the discretized \lc\ Schroedinger equation with
higher values of the cut-off $K$.
We hope that the goal of constructing a sensible interacting
non-critical string model with $c>1$ is not beyond reach.

\ack
We thank M. Martin-Delgado for advice on {\it Mathematica}.
I. R. K. is supported in part by
DOE grant DE-AC02-76WRO3072,
NSF Presidential Young Investigator Award PHY-9157482,
James S. McDonnell Foundation grant No. 91-48, and
an A. P. Sloan Foundation Research Fellowship. S. D. is supported by
S.E.R.C.(U.K.) post-doctoral fellowship RFO/B/91/9033.

\singlespace
\refout

\bigskip
\singlespace
\centerline{Figure Captions.}

Figure 1. -- The variation with $x$ of the lowest 16 eigenvalues (out of a
total of 19) for cutoff $K=7$.

Figure 2. -- The solid curves show the variation with $x$ of the mean
square separation between eigenvalues for the lower half of the states. The
total number of states for $K=9,10,11,12$ is $59,107,187,351$
respectively.
The dashed curve shows the variation with $x$ of the lowest eigenvalue in
each case.

Figure 3. -- The variation with $x$ of the difference between the lowest
two eigenvalues for $K=8,10,12$. The well becomes deeper with increasing $K$.
\bye